\documentclass[aps,11pt,preprintnumbers,axodraw]{revtex4}
\usepackage{epsfig}
\usepackage{amsmath}
\usepackage{bm}
\usepackage{times}
\usepackage{graphicx}
\usepackage{color}

\def\nn{\nonumber}
\def\bea{\begin{eqnarray}}
\def\eea{\end{eqnarray}}
\def\ba{\begin{eqnarray}}
\def\ea{\end{eqnarray}}
\def\be{\begin{equation}}
\def\ee{\end{equation}}

\begin{document}
\preprint{CALT 68-2759}
\title{\Large Scalar Representations and Minimal Flavor Violation}
\author{Jonathan M. Arnold$^{1}$            }

\author{Maxim Pospelov$^{2,3}$}

\author{Michael Trott$^{3}$}
\author{Mark B. Wise$^{1}$}

\affiliation{$^{1}$ California Institute of Technology, Pasadena, CA, 91125, USA,}

\affiliation{$^{2}$Department of Physics and Astronomy, University of Victoria BC, V8P-1A1, Canada, 
}

\affiliation{$^{3}$ Perimeter Institute for Theoretical Physics, Waterloo, ON N2J-2W9, Canada.}

\date{\today}
\begin{abstract}
We discuss the representations that new scalar degrees of freedom (beyond those in the minimal standard model) can have if they couple to quarks in a way that is consistent with minimal flavor violation. 
If the new scalars are singlets under the flavor group then they must be color singlets or color octets. In this paper we discuss the allowed representations and renormalizable couplings when the new scalars also transform under the flavor group.  We find that color ${\bar 3}$ and $6$ representations are also allowed. We focus on the cases where  the new scalars can have renormalizable Yukawa couplings to the quarks  without factors of the quark Yukawa matrices. The renormalizable couplings in the models we introduce automatically conserve baryon number.
\end{abstract}
\maketitle
\section{Introduction}

Theoretical and experimental evidence exists that hadron colliders will likely discover a scalar sector 
composed of effective or fundamental
scalar degrees of freedom.
The scalar sector could simply complete the Standard Model (SM) or it could be a low energy signature of a more UV insensitive theory.
When the SM is viewed as an effective theory, model building aimed at solving the hierarchy problem has frequently 
motivated the consideration of new scalars beyond the Higgs; including extra Higgs doublets and new colored 
scalars.

When one considers a scalar sector from the point of view of discovery potential at hadron colliders, new colored scalars 
with renormalizable couplings to standard model fields are particularly interesting. Such scalars can have 
large production cross sections in the $\rm pb$ range when their masses are low enough to be produced at the Tevatron and LHC.
Discovering these particles depends on SM background limitations and the couplings of the new states to the SM.

The couplings of new scalars to the SM quarks are significantly constrained from low energy experiments.
The consistency of the observed flavor changing neutral currents (FCNC) including $b \rightarrow s+\gamma$ and $K^0-{\bar K}^0$ mixing 
with the predictions of the SM, dictates that the effective couplings of these states to the quarks are very small due to a small coupling or large mass scale, or have a non-generic symmetry structure.
One way to allow new physics at the weak scale while being consistent with these constraints is if the new physics automatically is 
flavor conserving apart from the effects of the standard model Yukawa couplings. In the minimal standard model, the quarks couple to the Higgs doublet via the Yukawa terms,
\begin{equation}
{\cal L}_Y=g^{~~i}_{U~j}{\bar u}_{iR} Q_L^j H+g^{~~i}_{D~j}{\bar d}_{iR} Q_L^j H ^{\dagger}+{\rm h.c.}, 
\end{equation}
where color and $\rm SU(2)_L$ indices have been suppressed. These are the only terms in the standard model that break the $\rm SU(3)_{U_R}\times SU(3)_{D_R} \times SU(3)_{Q_L}$  flavor symmetry. Minimal flavor violation (MFV) \cite{Chivukula:1987py,Hall:1990ac,D'Ambrosio:2002ex} asserts  that any new physics (beyond that in the minimal standard model) also has this flavor symmetry only broken by insertions of the Yukawa coupling matrices, $g_U$ and $g_D$. Using the tadpole method, one can construct allowed new physics terms by endowing the Yukawa matrices with the transformation laws,
\begin{equation}
g_U \rightarrow V_U \, g_U \, V_Q^{\dagger},~~~~~~~~~~g_D \rightarrow V_D \, g_D \, V_Q^{\dagger},
\end{equation}
where $V_U$ is an element of $\rm SU(3)_{U_R}$, $V_D$ is an element of $\rm SU(3)_{D_R}$, and $V_Q$ is an element of $\rm SU(3)_{Q_L}$. Imposing MFV assures that flavor changing neutral currents induced from new physics at the weak scale will be acceptably small. MFV can be formulated up to linear order in top Yukawa insertions \cite{Chivukula:1987py,
Hall:1990ac,D'Ambrosio:2002ex,Cirigliano:2005ck,Buras:2003jf,Branco:2006hz},
or extended to a nonlinear representation of the symmetry
\cite{Feldmann:2008ja,Kagan:2009bn,Feldmann:2009dc}.

The Higgs doublet also couples to leptons. Assuming that there are no right handed neutrinos in the low energy effective theory, the small neutrino masses come from non-renormalizable dimension five operators. The renormalizable  Yukawa couplings of the Higgs doublet that give mass to the charged leptons are, 
\begin{equation}
\label{leptonyukawa}
{\cal L}_Y=g^{~~i}_{E~j}{\bar e}_{iR}L_L^j H^{\dagger}+{\rm h.c.},
\end{equation}
MFV can be extended to the lepton sector \cite{Cirigliano:2005ck} by enlarging the flavor group to, $\rm SU(3)_{U_R}\times SU(3)_{D_R} \times SU(3)_{Q_L}\times SU(3)_{E_R}\times SU(3)_{L_L}$.  Then Eq.~(\ref{leptonyukawa})
 is invariant uder the flavor group provided the lepton Yukawa coupling constant matrix $g_E$  is taken to transform as,
\begin{equation}
g_E \rightarrow V_E\, g_E \,V_L^{\dagger},
\end{equation}
where $V_E$ is an element of $SU(3)_E$ and $V_L$ is an element of $SU(3)_L$.

Assuming that any new scalars are singlets under the flavor group and have tree level quark couplings,  Manohar and Wise \cite{Manohar}  showed that the new scalars must have the same $\rm SU(2)\times U(1)_Y$ gauge quantum numbers as the Higgs doublet and can be either color singlets or color octets. The phenomenology of color octets has been considered elsewhere in some detail \cite{Manohar1,Dorsner:2007fy,Gresham:2007ri,Kim:2008bx,Idilbi:2009cc,Gerbush:2007fe,Burgess:2009wm,FileviezPerez:2008ib,Perez:2008ry,FileviezPerez:2009ud,Losada:2009yy}.  In this brief report,  we consider the generalization of this result to the case where the scalars transform under the flavor group. We restrict our attention to the cases where the new scalars can Yukawa couple to the quarks in a way that is invariant under the flavor group (with no insertions of $g_U$ and $g_D$).  
Scalars with couplings to SM fields similar to the ones we discuss sometimes go by the name `diquarks' and have been previously considered due to top down model building 
motivations \cite{Barbier:2004ez,Mohapatra:2007af,Chacko:1998td,Babu:2006xc,Babu:2008rq,Chen:2008hh} and due to their discoverability potential at the LHC \cite{DelNobile:2009st,Han:2009ya, Bauer:2009cc}.

For definiteness lets start with the case where the new scalars $S$ have tree level couplings to the diquark pair $u_R \, u_R$. The scalars $S$ must have hypercharge $Y=-4/3$ and be singlets under the weak $\rm SU(2)$ gauge group. They can be a color ${\bar 6}$ or $3$. In the case that they are in a color $3$ the diquark is antisymmetric under interchange of color and Lorentz spinor indices and so must be in a $3$ of $\rm SU(3)_{U_R}$ so that it is antisymmetric in flavor indices. If the new scalars are a color $\bar{6}$ the diquark is symmetric in the color indices, antisymmetric in Lorentz spinor indices and so it must be symmetric under interchange of flavor indices putting it in the ${\bar 6}$ representation of $\rm SU(3)_{U_R}$.  These two cases are respectively cases V and VI of Table1. 

In case V the  Lagrange density contains the term, $\eta \, \epsilon^{\alpha \beta \gamma} \,  \epsilon_{ijk} \, \left(u_{R\alpha}^i \, u_{R\beta}^j \right) \, S_{\gamma}^k +{\rm h.c.}$, where $\eta$ is a constant.  If MFV is not extended to the lepton sector then the Lagrange density can also contain the term, $\chi \left(d_{R\alpha}^i e_R^j \right)S^{* \alpha }_k \left(g_Ug_D^{\dagger}\right)^k_{~i}+{\rm h.c.}$, where $\chi$ is a constant. Hence, for case V MFV must be extended to the lepton sector for the  gauge symmetries and global flavor symmetry  to imply that the allowed renormalizable couplings  automatically conserve baryon number.

\begin{table}[tb]
\begin{center}
\begin{tabular}[t]{|c|c|c|c|c|c|}
  \hline
  \hline
   Case &$SU(3)_c$ & $SU(2)_{L}$ &$U(1)_Y$  & $SU(3)_{U_R} \times SU(3)_{D_R} \times SU(3)_{Q_L}$ & Couples to \\
  \hline
I &      1& 2& 1/2  & (3,1,${\bar 3}$) & ${\bar{u}_R}$ \, $Q_L$ \\
II &      8& 2& 1/2  & (3,1,${\bar 3}$) & ${\bar{u}_R}$ \, $Q_L$ \\
 III &         1& 2& -1/2  &(1,3,${\bar 3}$)& ${\bar{d}_R}$ \, $Q_L$ \\   
 IV &         8& 2& -1/2  &(1,3,${\bar 3}$)& ${\bar{d}_R}$ \, $Q_L$ \\   
 V&    3& 1& -4/3 & (3,1,1) & $u_R$ \, $u_R$ \\
   VI&  $\bar{6}$ &1& -4/3 & (${\bar 6}$,1,1) & $u_R$ \, $u_R$  \\
    VII&    3& 1& 2/3 & (1,3,1)  & $d_R$ \, $d_R$ \\
  VIII&    $\bar{6}$ &1& 2/3 & (1,${\bar 6}$,1)  & $d_R$ \, $d_R$  \\
   IX&  3 & 1& -1/3 & (${\bar 3}$,${\bar 3}$,1) & $d_R$ \, $u_R$ \\
   X&   $\bar{6}$ & 1& -1/3 & (${\bar 3}$,${\bar 3}$,1) & $d_R$ \, $u_R$ \\
    XI& 3  & 1 & -1/3 & (1,1,${\bar 6}$) & $Q_L$ \, $Q_L$ \\ 
    XII&  $\bar{6}$  & 1 & -1/3 & (1,1,3) & $Q_L$ \, $Q_L$ \\  
    XIII & 3&3&-1/3&(1,1,3)&$Q_L$ \, $Q_L$ \\  
   XIV&  $\bar{6}$  & 3 & -1/3 & (1,1,${\bar 6}$) & $Q_L$ \, $Q_L$ \\    
  \hline
  \hline
\end{tabular}
\end{center}
\caption{Different scalar representations that are not singlets under the flavor group that satisfy minimal flavor violation without the insertion of spurion Yukawa
fields that transform under the flavor symmetry.}
\label{int}
\end{table}

We are interested in models where the renormalizable couplings in the Lagrange density automatically conserve baryon number. For cases $\rm I - IV$ the new scalars have no baryon number and baryon number conservation is an automatic consequence of the gauge symmetries.  In cases $\rm  V-XIV$ the new scalars  have baryon number $-2/3$.  In the cases where the scalars transform under the ${\bar 6}$ representation the gauge symmetries alone imply automatic baryon number conservation. We also find that this is true in case VII. In cases V, IX , XI  and XIII,  if MFV is extended to the lepton sector, then the flavor and gauge symmetries combined restrict the renormalizable couplings to conserve baryon number.  For most of the remainder of this paper we focus on the cases where the scalars are weak SU(2) gauge singlets since that naturally suppresses the $\rm T$ precision electroweak parameter, i.e., cases V-XII. In those cases the scalars have baryon number  $-2/3$ . In the next section we discuss the Lagrange density for cases V-XII of Table 1 and some of the phenomenological consequences are discussed in Section III.

\section{Tree Level Lagrangians}
 Here we step through the construction of the tree level Lagrange density for cases V-XII of Table 1.  Before doing this we set some notation. We use primed fields for states that are mass eigenstates and unprimed fields for the weak eigenstates. So, $u^{\prime 1}=u$, $u^{\prime 2}=c$, $u^{\prime 3}=t$ and similarly for the down quark fields. The quark mass matrices are diagonalized by the transformations,
 \begin{equation}
 {\cal U}(u,R)^{\dagger} \, g_U \, {\cal U}(u,L)={\sqrt{2}\mathcal{M}_u \over v},~~~~~~{\cal U}(d,R)^{\dagger} \, g_D \, {\cal U}(d,L)={\sqrt{2}\mathcal{M}_d \over v}.
 \end{equation}
 Here $\mathcal{M}_u$ and $\mathcal{M}_d$ are the diagonal up-type and down-type quark mass matrices and $v$ is the vacuum expectation value of the neutral component of the Higgs doublet field. The weak eigenstate and mass eigenstate quark fields are related by,
 \begin{eqnarray}
 &&u_L={\cal U}(u,L) u'_L,~~~~~u_R={\cal U}(u,R) u'_R, \nonumber \\
 &&d_L={\cal U}(d,L) d'_L,~~~~~u_R={\cal U}(d,R) d'_R.
 \end{eqnarray}
 We will neglect all the quark masses except the top quark mass. 
 The Cabbibo-Kobayashi-Maskawa (CKM) matrix $V_{CKM}$ is related to the matrices that diagonalize the quark mass matrices by,
 \begin{equation}
 V_{CKM}={\cal U}(u,L)^{\dagger} \, {\cal U}(d,L).
 \end{equation}
 
 We will neglect all the quark masses except the top quark mass so we can set, ${\cal U}(d,R) ={\cal U}(d,L) ={\bf{I}}$ and we will adopt that approximation for the rest of this paper. The new scalar is denoted by $S$.
 \subsection{Case V}
 
 In this case the Yukawa coupling of the new scalar to the quarks takes the form,
 \begin{equation}
  \label{yukIII}
 {\cal L}_{YS}=\eta \, \epsilon^{\alpha \beta \gamma} \,  \epsilon_{ijk} \, \left(u_{R\alpha}^i \, u_{R\beta}^j \right) \, S_{\gamma}^k +{\rm h.c.},
 \end{equation}
 where we have explicitly displayed color and $\rm SU(3)_{U_R}$ flavor indices and suppressed the Lorentz spinor indices.
 Clearly it is convenient for going over to the quark mass eigenstate basis to define,
 $S ={\cal U}(u,R) S'$  so that in the quark mass eigenstate basis,
 \begin{equation}
 {\cal L}_{YS}=\eta \, \epsilon^{\alpha \beta \gamma} \, \epsilon_{ijk} \,  \left(u_{R\alpha}^{\prime i } \, u_{R\beta}^{\prime j}\right) \, S_{\gamma}^{\prime k} +{\rm h.c.}.
 \end{equation}
The flavor symmetry is broken by insertions of the Yukawa matrix $g_U$, which generate corrections to Eq.~(\ref{yukIII}).
Every contraction of an $\rm SU(3)_{U_R}$ flavor index in (\ref{yukIII}) allows for the insertion of $g_U \,g_U^{\dagger}$, which transforms under 
the flavor symmetry,  $g_U \,g_U^{\dagger} \rightarrow V_U \left(g_U \, g_U^{\dagger} \right)V_U^{\dagger}$. The leading order in insertions of the top mass Yukawa couplings gives a correction to Eq.~(\ref{yukIII}):
\bea
 \Delta{\cal L}_{YS} &=& {\tilde\eta_1} \, \epsilon^{\alpha \beta \gamma}  \, \epsilon_{imk} \, \left(u_{R\alpha}^{i} \, u_{R\beta}^{j} \right) \, S_{\gamma}^{k}  \, \left(g_U g_U^{\dagger}\right)^m_{~~j}+
{\tilde\eta_2} \, \epsilon^{\alpha \beta \gamma}  \, \epsilon_{ijm} \, \left(u_{R\alpha}^{i} \, u_{R\beta}^{j} \right) \, S_{\gamma}^{k}  \, \left(g_U g_U^{\dagger}\right)^m_{~~k}+
{\rm h.c}, \nn \\
 &=&\epsilon^{\alpha \beta \gamma}\epsilon_{imk} \left({2 \mathcal{M}_u^2 \over v^2}\right)^m_{~~j}\left({\tilde\eta_1} 
\left(u_{R\alpha}^{\prime i } \, u_{R\beta}^{\prime j}\right) \, S_{\gamma}^{\prime k} \, -
{\tilde\eta_2} 
\left(u_{R\alpha}^{\prime i } \, u_{R\beta}^{\prime k}\right) \, S_{\gamma}^{\prime j} \right)+ {\rm h.c.}.
\eea
 Since the only quark mass we treat as non-zero is the top quark mass this becomes,
 \begin{equation}
 \Delta{\cal L}_{YS}=-{\tilde  \eta_1} \, \left({2 m_t^2 \over v^2}\right) \, \left(u_R S^{\prime 2}-c_R S^{\prime 1}\right) \, t_R+
2{\tilde  \eta_2} \, \left({2 m_t^2 \over v^2}\right) \, S^{\prime 3} u_Rc_R +
{\rm h.c.},
 \end{equation}
 where now we have suppressed color indices in addition to the spinor Lorentz indices.
Note that the formalism for the consistent treatment of multiple insertions of the $g_U \,g_U^{\dagger}$ factors has been developed \cite{Feldmann:2008ja,Kagan:2009bn,Feldmann:2009dc} but is 
beyond the scope of this work.
 
Neglecting insertions of the quark mass matrices,  the most general renormalizable scalar potential for the Higgs doublet $H$ and colored scalar $S$ is,
 \begin{equation}
 V(H,S)={\lambda \over 4}\left( H^{\dagger} H -v^2/2 \right)^2+m_S^2 \left(S^{\prime j}_{\alpha}S^{\prime * \alpha}_j \right)+{\lambda_S \over 2} \left(S^{\prime j}_{\alpha}S^{\prime * \alpha}_j \right)^2+\lambda_{SH}  H^{\dagger} H\left(S^{\prime j}_{\alpha}S^{\prime * \alpha}_j \right).
 \end{equation}
 Without insertions of the top quark mass all the scalars are degenerate. Factors of $g_Ug_U^{\dagger}$ can also be inserted in the scalar potential leading to mass splittings amongst the 
new scalars.  The mass spectrum with one insertion of $g_Ug_U^{\dagger}$ is given by
\bea
 m_{S^{\prime 1}}^2 &=&  m_S^2 + \frac{\lambda_{SH}}{2} \, v^2, \nn \\
 m_{S^{\prime 2}}^2 &=&  m_S^2 + \frac{\lambda_{SH}}{2} \, v^2, \nn \\
 m_{S^{\prime 3}}^2 &=& m_S^2 + \frac{\lambda_{SH}}{2} \, v^2  + {\tilde m}_S^2 \, {2 m_t^2 \over v^2}.
\eea 
where $ {\tilde m}_S^2$ corresponds to the corrections to the mass and $\lambda_{SH}$ potential term due to  $g_Ug_U^{\dagger}$ insertions.

 \subsection{Case VI}
  In this case the Yukawa coupling of the new scalar to the quarks takes the form,
 \begin{equation}
  \label{yukIV}
 {\cal L}_{YS}=\eta \, \left(u_{R\alpha}^i \, u_{R\beta}^j \right) \, S_{ij}^{\alpha \beta} +{\rm h.c.},
 \end{equation}
where, $S_{jk}^{\alpha \beta}=S_{jk}^{\beta \alpha}=S_{kj}^{\alpha \beta}$. It is convenient for going over to the quark mass eigenstate basis to define,
 $S_{jk}={\cal U}(u,R)_j^{*~m}{\cal U}(u,R)_k^{*~n} S^{\prime}_{mn}$  so that in the quark mass eigenstate basis Eq.~(\ref{yukIV}) becomes,
 \begin{equation}
 {\cal L}_{YS}= \eta \, \left(u_{R\alpha}^{\prime i} \, u_{R\beta}^{\prime j}\right) \, S_{ij}^{\prime \alpha \beta} +{\rm h.c.}.
 \end{equation}
Again we can insert factors of the quark mass matrix that explicitly break the flavor symmetries. At order $m_t^2$ the correction to the Yukawa couplings are,
\begin{equation}
 \Delta{\cal L}_{YS}= {\tilde  \eta} \, \left( {2m_t^2 \over v^2}\right) \, \left(t_{R\alpha} \, u_{R\beta}^{\prime j}\right)  \, S_{3j}^{\prime \alpha \beta} +{\rm h.c.},
 \end{equation}
The scalar potential is very similar to case V. Without insertions of the quark mass matrix it is,
 \begin{equation}
 V(H,S)={\lambda \over 4}\left( H^{\dagger} H -v^2/2 \right)^2+m_S^2 \left(S^{\prime *jk}_{\alpha \beta}S^{\prime \alpha \beta}_{j k}\right)+{\lambda_S \over 2} \left(S^{\prime *jk}_{\alpha \beta}S^{\prime \alpha \beta}_{j k}\right)^2+\lambda_{SH}  H^{\dagger} H\left(S^{\prime *jk}_{\alpha \beta}S^{\prime \alpha \beta}_{j k}\right).
 \end{equation}
At first order $m_t^2$ in the top quark mass the components $S^{\prime}_{3j}$ are degenerate and split from the other components  $S^{\prime}_{ab}$, where $a,b ={1,2}$.
 \subsection{Cases VII and VIII}
 These cases are similar to cases V and VI once one replaces $u_R \rightarrow d_R$. The only difference is that there are no corrections from insertions of the top quark mass matrix
 and thus, in our approximation, the scalars are degenerate in mass.

\subsection{Cases IX and X}

We start with case X, where without insertions of the  quark mass matrices the Yukawa couplings are,
\begin{equation}
{\cal L}_{YS}= \eta \, \left(u_{R\alpha}^i \, d_{R\beta}^a\right) \, S^{\alpha \beta}_{i a}.
\end{equation}
Transforming to the basis, $u_{R\alpha}^{\prime i}$ and $S^{\prime \alpha \beta}_{ia}$ (where $S^{\alpha \beta}_{i a}= {\cal U}(u,R)_i^{*~j}S^{\prime \alpha \beta}_{j a}$) this becomes,
\begin{equation}
{\cal L}_{YS}= \eta \, \left(u_{R\alpha}^{ \prime i} \, d_{R\beta}^a \right) \, S^{\prime \alpha \beta}_{i a},
\end{equation}
in the quark mass eigenstate basis. Recall that since we are neglecting the down type quark masses $d^{\prime}=d$. At order $m_t^2$ the Yukawa couplings of $S$ receive the correction,
\begin{equation}
\Delta{\cal L}_{YS}={\tilde \eta} \, \left({2 m_t^2 \over v^2}\right) \, \left(t_{R\alpha} \, d_{R \beta}^a \right) \, S^{\prime \alpha \beta}_{3 a}.
\end{equation}
The scalar potential for the Higgs doublet and $S$ is,

\begin{equation}
 V(H,S)={\lambda \over 4}\left( H^{\dagger} H -v^2/2 \right)^2+m_S^2 \left(S^{\prime*ja}_{\alpha \beta}S^{\prime \alpha \beta}_{j a}\right)+{\lambda_S \over 2} \left(S^{\prime *ja}_{\alpha \beta}S^{\prime \alpha \beta}_{j a}\right)^2+\lambda_{SH}  H^{\dagger} H\left(S^{\prime *ja}_{\alpha \beta}S^{\prime \alpha \beta}_{j a}\right).
 \end{equation}

This potential receives corrections at order $m_t^2$ that split the mass of $S_{3a}^{\prime}$  from the other components.
The discussion for case IX is very similar.
\subsection{Case XI}

The leading Yukawa couplings are,
\bea
{\cal L}_{YS} &=& {\eta} \,  \epsilon^{\alpha \beta \gamma} \left(Q_{L \alpha}^{ i} \, Q_{L \beta}^{ j}\right) \, S_{\gamma i j } + {\rm h.c.}, \nn \\
 &=& 2 \, {\eta} \, \epsilon^{\alpha \beta \gamma} \left(u_{L \alpha}^ i \, d_{L \beta}^{j} \right) \, S_{\gamma i j} + {\rm h.c.},
\eea
where $S_{\gamma i j} = S_{\gamma j i}$.  As before, we write this in the mass eigenstate basis, using ${\cal U}(u,L)=V_{CKM}^{\dagger}$ and now defining $S_{i j} = {\cal U}_{i}^{*l}(u,L) \, {\cal U}_{j}^{*m}(u,L) \, S_{l m}^{\prime}$, we have,
\begin{equation}
{\cal L}_{YS} = 2 \,  \eta \, \epsilon^{\alpha \beta \gamma} \, \left(V_{CKM} \right)_{~~j}^{m} \, \left(u_{L\alpha}^{\prime  l} \,  d_{L \beta }^{ j} \right) \, S_{\gamma l m}^{\prime } + {\rm h.c.}
\end{equation}
At order $m_t^2$ the Yukawa couplings receive a correction,
\begin{equation}
\Delta {\cal L}_{YS} =  {\tilde \eta} \, \left(\frac{2 m_t^2}{v^2}\right) \,  \epsilon^{\alpha \beta \gamma} \left(V_{CKM}\right)^m_{~~j} \, \left(t_{L \alpha} \, d_{L \beta}^j \right) \,  S_{\gamma 3 m}^{\prime } + \textrm{h.c.}
\end{equation}

Ignoring insertions of the quark mass matrix, the scalar potential takes the form,
\begin{equation}
 V(H,S)=\frac{\lambda}{4} (H^{\dagger} H-v^2/2)^2+m_S^2 (S^{\prime *  \alpha i j} S_{\alpha i j}^{\prime }) +\frac{\lambda_S}{2}(S^{\prime *\alpha  i j} S_{ \alpha i j}^{\prime })^2+\lambda_{SH} H^{\dagger} H (S^{\prime * \alpha i j} S_{\alpha i j}^{\prime})
\end{equation}

\subsection{Case XII}

The leading Yukawa couplings are,
\bea
{\cal L}_{YS}&=& \eta\, \epsilon_{ijk}  \, Q_{L \alpha}^i \, Q_{L \beta}^j \, S^{\alpha \beta k}, \nn \\
&=& 2\, \eta \, \epsilon_{ijk} \,u_{L \alpha}^i \, d_{L \beta}^j \, S^{ \alpha \beta k}.
\eea
where,  $S^{k \alpha \beta}=S^{k  \beta \alpha}$. Going over to the quark mass eigenstate basis,  using ${\cal U}(u,L)=V_{CKM}^{\dagger}$ (since the weak and mass down type quarks coincide)  and defining $S^k= {\cal U}(u,L)^k_{~j }S^{\prime j}$, we find that,
\begin{equation}
{\cal L}_{YS}=2 \, \eta \, \epsilon_{i p m } \, \left(V_{CKM}\right)_{~~k}^p \, u_{L \alpha}^{\prime i}  \, d_{L\beta}^{k } \, S^{\prime \alpha \beta m}.
\end{equation}
At order $m_t^2$ the correction to the Yukawas is
\begin{equation}
\Delta {\cal L}_{YS} =  {\tilde  \eta} \, \left(\frac{2 m_t^2}{v^2}\right) \,  t_{L \alpha} \left( \left(V_{CKM}\right)_{~~k}^{1} \, d_{L \beta}^k \, S^{\prime  \alpha \beta 2} - \left(V_{CKM}\right)_{~~k}^{2} \, d_{L\beta}^k \, S^{\prime  \alpha \beta 1} \right) + {\rm h.c.}
\end{equation}

At leading order the scalar potential is
\begin{equation}
V(H,S)=\frac{\lambda}{4} (H^{\dagger} H-v^2/2)^2+m_S^2 (S_{\alpha \beta i}^{\prime * } S^{\prime \alpha \beta i})+\frac{\lambda_S}{2}(S_{\alpha \beta i}^{\prime * } S^{\prime \alpha \beta i})^2+\lambda_{SH} H^{\dagger} H (S_{\alpha \beta i}^{\prime *} S^{\prime \alpha \beta i}).
\end{equation}

\section{Phenomenological Aspects}

\subsection{Electroweak Precision Data and LEPII constraints}
Constraints on the allowed $(m_S,\eta)$ come from Tevatron data, LEPII direct production bounds, EWPD and low energy experiments
(flavor physics, EDM's). We will discuss each of these constraints in turn.
We begin with EWPD bounds as they are particularly concise and free of unknown parameter dependence as the gauge interactions of these scalars are fixed
and these bounds are not dependent on backgrounds and cuts.

\begin{figure}[hbtp]
\centerline{\scalebox{1.3}{\epsfig{file=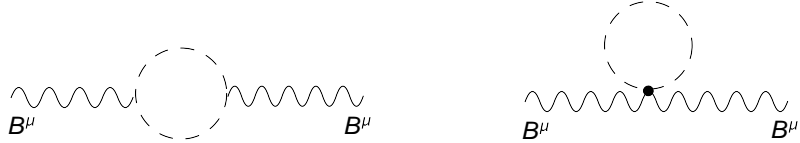}}}
\caption{The oblique corrections of the colored scalars of cases V thru XII.} 
\end{figure}

Consider the case that the Yukawa coupling constant, $\eta$ is  small enough that these scalars make only a small one loop contribution to the $Z$ decay rates to quarks. 
Then  the oblique corrections  \cite{Holdom:1990tc,Peskin:1991sw,Altarelli:1990zd} can provide a lower bound on the masses. As mentioned earlier we are only considering cases V-XII of Table 1 so the new scalars are SU(2) singlets.  Since the scalars do not couple to the  SU(2) gauge bosons, for $m_S^2 \gg M_W^2$, one can consider only the oblique parameter $Y$ defined by $Y= \, g_1^2 \, M_W^2 \, \Pi^{''}_{BB}(0)/2$ \cite{Barbieri:2004qk}, where $ \Pi_{BB}$ the hypercharge gauge boson self energy and $g_1$ is the weak hypercharge . We find that,
\begin{equation}
\label{y}
Y=-\left({M_W^2 \over m_S^2}\right){\alpha_{\rm em} \, Y_S^2 \, d(r_c) \, d(r_F) \over 120 \, {\rm cos}^2\theta_W  \, \pi}
\end{equation}
where $d(r_c)$ is the dimension of the color representation the new scalar is in, $d(r_F)$ is the dimension of the flavor representation, $\theta_W$ is the Weinberg angle and $\alpha_{\rm em} \simeq 1/128$ is the electromagnetic fine structure constant. To see how stringent a constraint the experimental value $Y=(4.2 \pm 4.9 )\times 10^{-3}$  \cite{Barbieri:2004qk} places on $m_S$ we consider case VI of Table 1. There $Y_S=4/3$,  $d(r_c)=d(r_F)=6$. Then Eq.~({\ref y}) gives, $Y\simeq -10^{-3}(M_W/m_S)^2 $ so a scalar mass of a hundred ${\rm GeV}$ gives a contribution to $Y$ that is less than the experimental error.

To obtain a precise lower bound on $m_S$ from electroweak precision data (EWPD) one can use the
6-parameter description ($\rm STUVWX$), that does not expand in $q^2$ \cite{Burgess:1993mg,
Maksymyk:1993zm,Burgess:2009wm}. This formalism is not limited to the case $m_S^2 \gg M_W^2$ and in general, the $\rm STUVWX$ formalism reduces
to the three-parameter STU (and its extension to include the parameter $Y$ above) when all new particles become very heavy compared to the massive gauge bosons. 
The lower bound on the masses $m_S$ of these states from this analysis are very weak since  there is no custodial symmetry ($SU(2)_c$) violation. 
In cases V-XII precision electroweak physics does not constrain $m_S$ to be above $100{\rm GeV}$ at the $95\%$ level.  At the $68\%$ level, only case VI is constrained to be above $100{\rm GeV}$, with a lower bound of $121{\rm GeV}$.
Note that new scalars that are a $3$ or a $2$ under $\rm SU(2)_L$ will be significantly more constrained due to $\rm SU(2)_c$ violation. 
It is still possible that such  states can be light enough to be discovered at LHC if approximate custodial symmetry is present in the 
scalar potential.

However, if the mass of the scalars were
smaller than approximately half the final operating energy at LEPII ($209 \, {\rm GeV}$) the scalars would have been produced centrally at LEPII  and decayed to 
give anomalous 2 and 4 jet final states.  No such excesses were observed and thus a kinematic lower bound of $105 \, {\rm GeV}$ for these scalars is imposed by 
LEPII direct production, which is stronger than the EWPD constraints in most cases.


\subsection{Tevatron Physics}

Stronger bounds can come from direct production constraints at the Tevatron, although these bounds will depend in many cases on the 
unknown parameters in the model and be limited by SM backgrounds. The most direct bounds are on scalars that are light enough to be produced 
in the s channel (corresponding to the cases V-XII we are considering)
that decay to  $u \, u$ and $u \, c$, $b \, u$ etc. The relevant final states at hadron colliders for the various cases are given in Table II.

\begin{table}[tb]
\begin{center}
\begin{tabular}[t]{|c|c|}
  \hline
  \hline
   Case & s-channel Final States \\
  \hline
   V& $S \rightarrow(t\,j[u,c],\, j[c,u]  \, j[u,c] )$ \\
    VI&  $S \rightarrow (t\,t,\,t\,j[u,c],\,jj[uu, cc, uc])$ \\
  VII&  $S \rightarrow (b \, j[d, s], j[s,d] \, j[d, s]) $  \\
   VIII& $S \rightarrow (b \, b, \,bj[d,s],\, j j[dd,ss,ds])$ \\
      IX & $S \rightarrow (t \, j[d,s,b],  b \, j[u,c],  j \,j[ud,us,cd,cs])$ \\
    X & $S \rightarrow (t \, j[d,s,b],  b \, j[u,c],  j \,j[ud,us,cd,cs])$  \\
    XI &   $S \rightarrow (t \, j[d,s,b],  b \, j[u,c],  j \,j[ud,us,cd,cs])$  \\
    XII &  $S \rightarrow (t \, j[d,s],  b \, j[u,c],  j \,j[us,cd])$  \\
  \hline
  \hline
\end{tabular}
\end{center}
\caption{Final state signatures for the cases V-XII for the Yukawa decay of the new scalars when they are produced in the $s$ channel. 
A $b$ or $t$ initiated jet is directly labelled and the quark states that initiate jets without a heavy flavor tag are labeled in the brackets.
The new scalars can also have $s$ channel production and decays as above with an extra gluon initiated jet in the final state radiated off the
initial state quarks or the new scalar. These final states are included in the NLO analysis of the production of new scalars of this type \cite{Han:2009ya}.
Note that for cases XI and XII the relative number of final state fermions is ordered by CKM suppression while for cases IX and X the same number of 
final state $jj$ events (for example) is expected for the various quarks. This has the potential to discriminate between the various cases.} 
\label{int}
\end{table}

The production of new scalars of this form has been studied in the past \cite{Mohapatra:2007af} and NLO QCD corrections have been examined \cite{Han:2009ya}.
These states could be detected as new resonances in the dijet invariant mass spectrum measured at Tevatron despite the enormous QCD background.
Consider $m_S =  1 \, {\rm TeV}$, then dijet resonance searches \cite{Aaltonen:2008dn} require that the new scalar production cross section is such that
\bea
\eta^2 \, \hat{\sigma}(m_S = 1 \, {\rm TeV})\times Br(S \rightarrow jj)\times \mathcal{A}(|y| > 1) \lesssim  10^{-1} \, {\rm pb}
\eea
at $95 \, \%$ confidence level where $\hat{\sigma} = \sigma/\eta^2$. 
We will assume $Br(S \rightarrow jj) \sim 1$. $\mathcal{A}(|y| > 1)$ is the acceptance cut on the rapidity of the produced dijets.
 Setting order one factors to unity (e.g., $\mathcal{A}(|y| > 1)$ and $Br(S \rightarrow jj)$)
the dijet resonance constraint on $\eta^2 \, \hat{\sigma}$ is
\bea
\eta^2 \, \hat{\sigma}(1 \, {\rm TeV}) \lesssim  10^{-1} \, {\rm pb}.
\eea
For comparison, the cross section values of \cite{Han:2009ya} for $m_S = 1 \, {\rm TeV}$ for various initial state quarks range from $10^{-1} {\rm pb} \, [uu]$ to $10^{-4} \, {\rm pb} \, [ss]$.
Sextet diquarks that couple at tree level to $uu$  (case VI) has the largest cross section at NLO  $\hat{\sigma} = 10^{-1} \, {\rm pb}$ \cite{Han:2009ya} but these 
still allow $\eta^2 \sim 1$. Repeating this analysis for $m_S =  500 \, {\rm GeV}$ one finds
\bea
\eta^2 \, \hat{\sigma}(500 \, {\rm GeV}) \lesssim  1 \, {\rm pb}.
\eea
For this lower mass value, the most stringent cases are VI which requires $\eta \lesssim 0.1$ (for $uu$ initial states)
and cases IX-XII which require $\eta \lesssim 0.1$ (from $du$ initial states). Thus we take a conservative lower bound in this case of 
$m_S >  500 \, {\rm GeV}$ for these states when $\eta \gtrsim 0.1$ from dijet resonance constraints.

The angular distribution of the cross section for the diquark does not depend on the polar angle from the beam
as it is a spinless resonance. 
Conversely there exists an approximate Rutherford scattering angular dependence for QCD dijet events
in the polar angle. Translating limits on deviations in the angular distributions of dijet events \cite{:2009mh,Abe:1996mj} into 
a bound on diquarks is a promising approach to constrain $m_S$ and $\eta$, although model independent exclusion bounds are not presently available.
To maximize the possibility of discovering diquark events, resonance searches and angular distribution studies should be performed 
minimizing the QCD background with a rapidity cut $y \ll 1$.



The heavy $S$ scalars are also produced in pairs and their subsequent decay to jets leads to four jet events.  This production mechanism is independent of the coupling $\eta$ to quarks and will be the dominant production mechanism for very small $\eta$.  A particularly attractive case is model VIII where pair production of $S_{33} S^*_{33}$ leads to $bb \bar b\bar b$ four jet final states. It is likely that an $S_{33}$ with a mass less than about $300{\rm GeV}$ could be discovered in this channel at the Tevatron. For a discussion of this in the case of color octet scalars see \cite{Gerbush:2007fe}.

Searches for scalars that lead to final states with two top quarks,  for example via $u+g \rightarrow S t \rightarrow tt\bar u$  were performed in \cite{Aaltonen:2008hx}. Model VI can lead to $t \bar t u$ final states  via the parton level production mechanism $g u \rightarrow S {\bar t } \rightarrow  {\bar t} tu$ however, final states of the type $tt{\bar u}$  do not arise from the models we consider.

For physics at energy scales much less than $m_S$ there are four quark operators induced when this state is integrated out.  We  consider here operators that are invariant under the flavor symmetries 
and arise at tree level. The coefficients of these operators are restricted by analysis of hadronic final states at the Tevatron. For definiteness consider cases V and VI. Integrating out the scalar in case V gives the effective Lagrangian,
\bea
\Delta{\cal L}_{\rm eff}=2{\eta^2 \over m_S^2}\left(u_{R\alpha}^{\prime i} u_{R\beta}^{\prime j} \right)\left({\bar u}_{R i}^{\prime \alpha} {\bar u}_{Rj}^{\prime \beta}-{\bar u}_{R i}^{\prime \beta} {\bar u}_{Rj}^{\prime \alpha}\right),
\eea
while the induced four quark operator in case VI  is 
\bea
\Delta{\cal L}_{\rm eff}={1 \over 2} \frac{\eta^2}{m_S^2} \left(u_{R\alpha}^{\prime i} u_{R\beta}^{\prime j} \right)\left({\bar u}_{R i}^{\prime \alpha} {\bar u}_{Rj}^{\prime \beta}+{\bar u}_{R i}^{\prime \beta} {\bar u}_{Rj}^{\prime \alpha}\right).
\eea
The other cases are similar to this example and the four quark  operators can be Fierz transformed into 
a more standard form. 
As the tree level exchange is proportional to $\eta^2$, constraints on four quark operators can only be 
translated into a bound on $\left(m_S,\eta \right)$ not affording a lower mass bound. However, another four quark operator is induced that is not dependent on an unknown coupling.
It comes from the gluon self energy bubble and the equations of motion are used to convert the operator with gluons into one with quarks. 
This contribution to the effective Lagrangian is,
\begin{equation}\label{effop}
\Delta{\cal L}_{\rm eff}={\alpha_s^2(m_S) \over 30 \, m_S^2} \, C(r_c) \, d(R_F) \, \left({\bar q} T^A \gamma^{\mu}q\right) \left({\bar q} T^A \gamma_{\mu}q\right),
\end{equation}
where $\alpha_s(m_S)$ is the running strong interaction fine structure constant evaluated at a subtraction point equal to $m_S$. 
Several bounds on four quark operators with the above forms are quoted in the literature  \cite{Lillie:2007hd, Cho:1993eu,shutalk}, however they are quite weak.   None of them  restrict $m_S$ to be above $100~{\rm GeV}$ for $\eta <1$. For light masses comparable to the energy used to probe
the four quark operators $m_S \sim E$, as is appropriate for Tevatron studies with light scalar masses, treating the effective dimension six operator as a contact interaction is not well justified.

The challenge of strongly constraining these models at the Tevatron, illustrated by the mass limits we have discussed, might lead one to believe that 
it is also very difficult to constrain these models at the LHC, but this is incorrect.
It has recently been emphasized \cite{Bauer:2009cc} 
that the relative parton luminosity at LHC compared to Tevatron for $uu$ initial states is such that for cases V,VI an early LHC run with $10 \, {\rm pb^{-1}}$ 
of data will have a greater sensitivity to these diquark models than Tevatron with $10 \, {\rm fb^{-1}}$ of data for large invariant mass states $m_S \sim 1-2 \, {\rm TeV}$,
making these states very interesting candidates for early LHC discovery in this mass range. 

It is interesting to note that scalars that are a $\bar{6}$ or a $3$ under $\rm SU(3)_c$ that allow couplings between $u$ and $t$, such as our models V,VI,
have been identified as promising candidates to explain the $\sim 2 \, \sigma$ anomaly measured in the top quark forward backward asymmetry through a $t$ channel 
exchange of the new scalar \cite{Shu:2009xf}.
 
\subsection{ Higgs Physics}

New colored scalars of this form can be consistent with EWPD and flavor physics and have the potential to modify the 
properties of the Higgs as inferred from EWPD. Performing a joint fit of the Higgs and the new scalars at one loop in the EWPD
one obtains the results in Figure 2. Here we have assumed that $\eta$ is very small. 

\begin{figure}[hbtp]
\centerline{\scalebox{0.48}{\epsfig{file=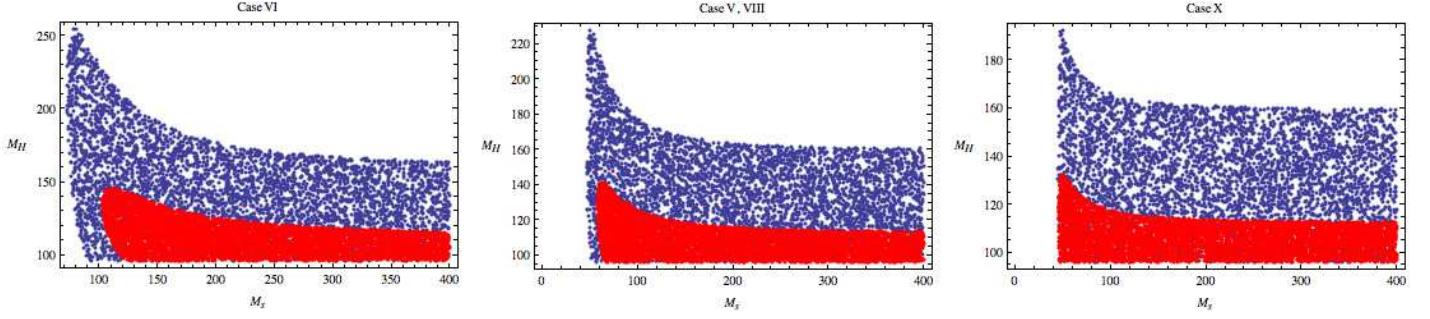}}}
\caption{The one loop joint fits of cases $V,VI,VIII,X$ and the Higgs mass (that is floated at one loop in the EWPD according to the procedure described in \cite{Burgess:2009wm}).
The red (solid grey) region is the $68 \%$ confidence region defined through the cumulative distribution function while the blue (dotted black) region is the $95 \%$ confidence region.
Only case VI allows  the fitted value of the Higgs mass to be raised significantly when the LEPII direct production bound of $m_S > 105 \, {\rm GeV}$ is imposed. 
The cases VII,IX not shown have a negligible impact on the fitted value of the Higgs mass at one loop. More dramatic effects on the 
fitted value of the Higgs mass are possible for cases I-IV and XIII-XIV due to the possibility of positive contributions of $\Delta {\rm T}$. } 
\end{figure}

In this case however, the data allows a scenario where the Higgs is relatively heavy while the new colored scalars can be relatively light.
When a new colored scalar exists and $m_S < m_h/2$ the Higgs branching ratio to easily detected
final states $\gamma \, \gamma$ and lepton signatures through $W^+ \, W^-$ can be suppressed due to the total decay width being modified, similar to the scenario in
\cite{Mantry:2007ar}. However, for most of the parameter space $m_S \gtrsim m_h$ and the Higgs decay properties are only modified
through the effects of integrating out the new scalar.

If the new scalars have masses less than a ${\rm TeV}$ then they can have a significant impact on the production rate for the Higgs boson at the LHC and the branching ratio for its decay to two photons.  Integrating out the scalars and using the notation of  \cite{Manohar1} the relevent terms in the effective Lagrangian that are relevant in these models are (For models V-XII, $ c_{WB}=c_W=0$.),
\begin{equation}
\Delta{\cal L}_{\rm Higgs}=- {c_G \over 2 \Lambda^2}g_s^2 H^{\dagger}H G_{\mu \nu}^A G^{A \mu \nu}-{c_B \over 2 \Lambda^2}g_1^2 H^{\dagger}H B_{\mu \nu} B^{ \mu \nu}
\end{equation}
where $\Lambda$ is a parameter with dimensions of mass. We find that,
\begin{equation}
{c_G \over \Lambda^2}=-{\lambda_{SH} d(r_F)C(r_c) \over 384 \pi^2 m_S^2},~~~~~{c_B \over \Lambda^2}=-{\lambda_{SH} Y_S^2d(r_F)d(r_c) \over 384 \pi^2 m_S^2},
\end{equation}
where ${\rm Tr}T^AT^B=C(r_c)\delta^{AB}$. For the representations we need:  $C(3)=1/2$, $C(6)=5/2$, and $C(8)=3$. To get a feeling for how big $c_G$ and $c_B$ might be lets take $\Lambda=1 \, {\rm TeV}$, $m_S=0.5 \, {\rm TeV}$ and $\lambda_{SH}=1$ in Case VI of Table 1. This gives, $c_G\simeq-0.016 $ and $c_B \simeq -0.068$ resulting in about a $20 \%$  decrease in the branching ratio of the Higgs boson to two photons and increases its production rate by about $40\%$. 

It is also possible that diquarks could be a source of Higgs production at LHC as diquarks
can emit single or pairs of Higgs particles with the rate depending on the value of $\lambda_{SH}$ in the scalar potential. However, a simple estimate of the rate of these emissions can be obtained by using a soft Higgs approximation where the scalar diquark energy and momenta are large compared to an emitted nonrelativistic Higgs of velocity ${\bf v}$. This
yields an estimate of the cross section for single Higgs production off singly produced diquarks  of order $\sigma \, \lambda_{SH}^2 (v/m_S)^2 \,  |{\bf v}|^3/(16 \pi^2) \sim 10^{-7} \, \sigma \, \lambda_{SH}^2$, where we have taken $m_S = 1 \, {\rm TeV}$ and $|{\bf v}| = 0.1$ and $\sigma$ is the cross section for diquark production. Considering the production cross sections of $1 \,{\rm TeV}$ diquarks  satisfy  $\sigma \lesssim \eta^2 \, 10^3 \,{\rm pb}$ \cite{Han:2009ya} at NLO this source of Higgs production is subdominant to gluon fusion for perturbative couplings.

\subsection{Low energy effects: Baryon Number Violation, Flavour Physics and Parity Violation}
 For each particular case in Table 1 baryon number violation by renormalizable interactions can be forbidden by gauge invariance and MFV. However if we allow fields corresponding to more than one case in the Table to occur this is no longer true. Consider, for example, combining cases V and VII. Then we have fields, $S_{V}$ and $S_{VII}$ with hypercharges -4/3 and 2/3 respectively. Renormalizable baryon number violating  Yukawa couplings of these scalars are forbidden by the combination of the gauge symmetries and MFV (extended to the lepton sector) however the following term in the scalar potential violates baryon number: 
 \begin{equation}
 \label{bnv}
 {\cal L}_{\rm bnv}=g_{\rm bnv} \epsilon^{\alpha \beta \gamma}\epsilon_{abc} S_{VII \beta}^{a} S_{VII \gamma}^{b} S^{j}_{V \alpha} \left(g_D g_U^{\dagger}\right)^c_{~~j}.
 \end{equation}
Here we did not set the light quark masses to zero. If we had, then the MFV would have forbidden this baryon number violating term.
Integrating out the $S$-fields, at low energy, one can see that a $\Delta b =2$ six-quark dimension 9 operator is induced. Assuming similar 
size $\eta$ couplings to all diquark combinations and a similar mass scale for all colored scalars, 
one can estimate the size of the corresponding coefficient as $G_{\rm bnv} \sim \hat{g}_{\rm bnv} \eta^3 m_u m_d v^{-2} m_S^{-5}$, where $\hat{g}_{\rm bnv}=g_{\rm bnv}/m_S$ is a dimensionless coupling. Such operators are capable of 
inducing nucleon-nucleon annihilation inside a nucleus, $A\to A-2~+[{\rm pions}]$, as well as neutron-antineutron oscillations. 
Experimental tests of the latter are sensitive to $G_{\rm bnv} \sim 10^{-25} ~{\rm GeV}^{-5}$ \cite{Babu:2008rq}. However, given a rather strong power dependence
on all the parameters involved, one cannot rule out a possibility of $m_S \lesssim $ TeV even in the presence of operator (\ref{bnv}). 

By design, a model with MFV is going to be less constrained by flavor physics than a model 
with an arbitrary flavor structure of $\eta$ couplings. Having said that, we cannot rule out some 
sizable flavor-changing effects with scalars coupled to the left-handed quarks. (For earlier discussion of 
flavor changing transition induced by elementary diquarks see \cite{Voloshin:2000uj}). For example, in case XIV
of Table 1, we can notice that the mass matrix of the scalars may receive a flavor-nondiagonal 
correction, quartic in the Yukawa coupling, 
\begin{equation}
\Delta V(S) = {\tilde m_S^2}   S^{*ij} \left( g_U^\dagger g_U \right)^{r}_{~~i} \left( g_U^\dagger g_U \right)^{s}_{~~j} S_{rs},
\end{equation}
where we have suppresed all indices but flavor $\rm SU(3)_{Q_L}$. 
Treating ${\tilde m_S^2}$ as a mass insertion, we integrate out $S$ fields to obtain 
the four-fermion operators, which now contain $\Delta F =2$ transitions. In particular,
the operator that induces $\Delta m_{B_d}$ is given by 
\begin{equation}
\label{FCNC}
{\cal O}_{\Delta B = 2} = \frac{|\eta|^2{\tilde m_S^2}}{m_S^4} ~ \frac{4m_t^4V_{tb}^{*2}V_{td}^2}{v^4}~(b_L b_L) ({\bar d}_L {\bar d}_L)
\end{equation}
After Fiertz transformation this operator will reduce to a familiar $(\bar d_L \gamma_\mu b_L)(\bar d_L \gamma_\mu b_L)$ combination. 
Since the mass splitting in the $B_d$ system agrees well with the SM, we have to require that the size of (\ref{FCNC}) 
is substantially less than that of the SM-induced operator. Given that ({\ref{FCNC}) does not contain any additional CKM or Yukawa suppression 
relative to the SM contribution, and that the SM amplitude is induced at a loop level, one has to conclude that 
in this particular case flavor physics provide sensitivity to 
\be
\frac{4 \, |\eta|^2 \, {\tilde m_S^2 \, m_t^4}}{v^2 \, m_S^4} \sim 10^{-4},
\ee
which can be easily satisfied with a modest hierarchy of $\tilde m_S$ and $m_S$, without requiring that the overall mass 
scale be so heavy as to rule out discovery at LHC. 

It is easy to see that the simple exchange of the colored scalars does not lead to the generation of the 
electric dipole moments (EDMs). Indeed, the EDMs are induced by the flavor-diagonal four-fermion operators
of the $S\times P$ type, $(\bar q q)(\bar q i \gamma_5 q)$ (see, {\em e.g.} \cite{Pospelov:2005pr}), while the diquark type 
scalar exchange is 
connecting particles of the same chirality, thus leading to $V\times V$, $A\times A$ and $V\times A$ Lorentz structures,
none of which is CP-violating with one flavor. Therefore, without mixing different types of colored scalars, only cases 
I-IV in the Table I could lead to EDM-inducing CP-odd operators. However, the MFV rule also guarantees that the 
chirality flip in the scalar-light quark vertex comes at the expense of a light quark Yukawa coupling in each vertex $\sim m_q/v$, 
which suppresses the induced EDMs to a ``safe'' level. The interaction with heavy quarks may induce EDMs at two-loop
level \cite{Weinberg:1989dx,Barr:1990vd}, but this also does not provide strong sensitivity to the scalar mass scale.  
The induced $V\times A$ operators, $(\bar q \gamma_\mu q)(\bar q \gamma_\mu\gamma_5 q)$,
generate $O(\eta^2m_S^{-2})$ corrections to the nucleon-nucleon parity violating interactions. 
However, there are problems even with the SM interpretation of the current nucleon spin dependent 
P-odd observables \cite{Haxton:2001ay}, and $\eta^2m_S^{-2} \sim G_F$ cannot be ruled out.

Although in our paper we concentrated on MFV, the alternative route of having very small Yukawa couplings of the new colored scalars $S$ is also possible for some of the models  and can have interesting phenomenological consequences. Diquark models VI, VII, VIII, X and XII have automatic baryon number conservation even if the couplings to leptons are not restricted by MFV.  For models without MFV the strongest constraint on the size of their Yukawa couplings is for scalars that  couple to $d_R$ or $Q_L$ quarks since tree level exchange of such scalars leads, for example, to operators of the type $({\bar d}_R {\bar d}_R)(s_R s_R)$ that contribute to the CP violation parameter $\epsilon_K$ of the kaon system. Denoting the typical size of a Yukawa coupling of the new scalars to the quarks by $\eta$ the contribution of such scalars to the CP violation parameter $\epsilon_K$ is of order, $|\Delta \epsilon_K| \sim \eta^2 10^{10}(100~{\rm GeV}/m_S)^2$. Demanding that this contribution  be an order of magnitude less than the measured value of $|\epsilon_K|$ gives the bound, $\eta <10^{-7}(m_S/100~{\rm GeV})$. Even if $\eta$ is very small such scalars are pair produced at the LHC with cross sections that are comparable to the $t \bar t$ production cross section if the scalar masses are around the top quark mass. Their lifetime is of order $\tau_S \sim \eta^{-2}(4\pi/m_S)$. For $m_S \sim 100{\rm GeV}$, a value, $\eta \sim 10^{-8}$, leads to displaced vertices with $c \tau > 1{\rm cm}$, and even smaller couplings may result in observable massive charged tracks within the detector. 

\section{Summary}
We have considered the generic consequences of new scalar particles with the quantum numbers of 
diquarks. We choose MFV as a guiding principle and have classified all possible types of the 
scalars that couple to quark bi-linears without insertions of Yukawa matrices at tree level. The assignment of quantum numbers under flavor 
transformations to the scalars allows one to drastically reduce the number of unknown couplings 
to different flavor combinations of the diquarks and improves the predictivity of the various models. The approximate preservation of flavor symmetries enables one to have
masses of these scalars $\lesssim {\rm TeV}$, which afford large cross sections, making these states very discoverable in early LHC data or at the Tevatron. 
Gauge symmetry and enforcing MFV forbids baryon number and flavor violating interactions,
and for many of the models gauge symmetry alone is enough to forbid baryon number violation by renormalizable interactions. In the models where gauge symmetry alone is enough to enforce baryon number conservation for the renormalizable interactions MFV can be relaxed and the new scalars can still be quite light. The most direct constraints on the models come from direct production at LEPII indicating $m_S > 105 \, {\rm GeV}$ and Tevatron data on resonance searches
where $m_S > 500 \, {\rm GeV}$ for $\eta > 0.1$.
There is great promise for increasing the bounds on colored scalars at the LHC and reach for a discovery of these states, even with the 
initial luminosity-limited data set. 

\subsection*{Acknowledgment}
We thank Pavel  Fileviez Perez for helpful comments on baryon number violation in the models we introduce. The work of M.B.W. was supported in part by the U.S. Department of Energy contract No. DE-FG02-92ER40171.  
M.B.W. is grateful for the hospitality of the Perimeter Institute where his participation in this work began.  The work of MP was supported in part by NSERC, Canada. Research at the Perimeter Institute
is also supported in part by the Government of Canada through NSERC and by the Province
of Ontario through MEDT. 



\end{document}